\documentclass[aps,pre,preprint,showpacs,amsmath,amsfonts,preprintnumbers]{revtex4}
\usepackage{epsfig}
\newcommand{\Op}[1]{{{\mathrm{\hat{#1}}}}}

\begin{document}
\title{Negativity as a distance from a separable state.}
\author{M. Khasin, R. Kosloff and D. Steinitz}
\affiliation{Fritz Haber Research Center for Molecular Dynamics, 
Hebrew University of Jerusalem, Jerusalem 91904, Israel}
\date{\today }

\begin{abstract} The computable measure of the mixed-state entanglement, the negativity, is shown to admit a  clear geometrical interpretation, when applied to  Schmidt-correlated  (SC) states: the negativity of a SC state   equals a distance of the state from a pertinent separable state. As a consequence, a
SC  state is separable if and only if its negativity vanishes. Another remarkable consequence is that the negativity of a SC can be estimated "at a glance" on the density matrix.  These results are generalized to  mixtures of SC states, which emerge in certain quantum-dynamical settings. 

\end{abstract}
\maketitle

Non local quantum correlations are a key resource in quantum information processing \cite{nielsen}. 
The exclusive quantum part of this
correlation has been termed entanglement. For  a pure bipartite state the extent of entanglement is well defined by  the 
Schmidt rank of the state \cite{schmidt,peres}, counting the number of non vanishing terms in the product states decomposition \cite{Virmani}. 
For mixed quantum states the notion of entanglement is more involved and various measures have been suggested \cite{Virmani}, each focusing on  particular aspects of this phenomenon.
In order to take advantage of the insights learned from different measures, it is advantageous to seek for classes of states where different entanglement measures agree \cite{Vollbrecht}. Such a class is the bipartite Schmidt correlated states  class   \cite{Rains,Rains_error, Virmani_sacchi}, where it has been shown that the distillable entanglement  \cite{Bennett_Wooters} and relative entropy of entanglement \cite{Vedral} both coincide and can be calculated by a simple formula
 \cite{Rains,Rains_error,Hiroshima, Vedral}.
The Schmidt-correlated (SC) states are defined as mixtures of pure states, sharing the same Schmidt bases  \cite{Rains,Rains_error, Virmani_sacchi}.
Such states naturally appear in a  bipartite system dynamics with additive integrals of motion (see below and Ref.\cite{Khasin06}). Hence, these states form an important class of mixed states from a quantum dynamical perspective.

\begin{figure}[t]
\epsfig{file=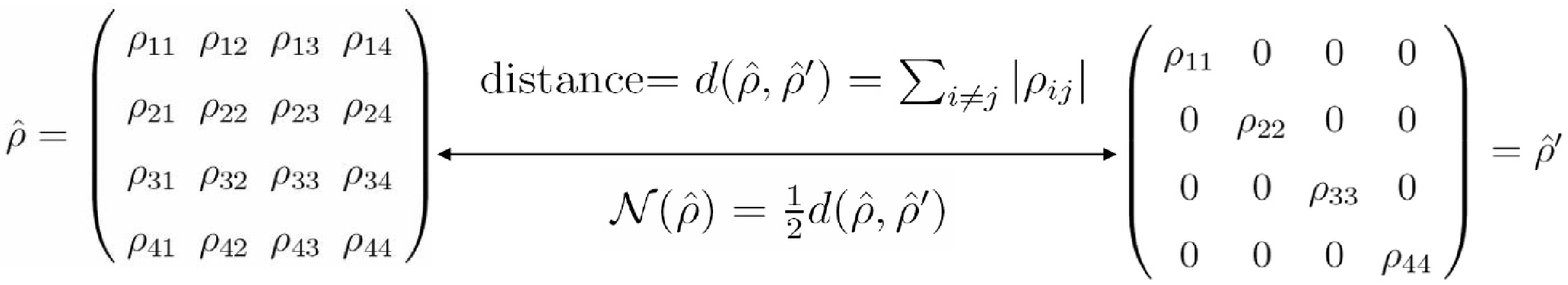, width=15.0cm, clip=} 
\caption{The negativity of a SC state $\Op \rho$ equals half the sum of the absolute values of the off-diagonal elements of the density matrix, which is a distance of the state state from the separable state $\Op \rho'$.}
\label{one}
\end{figure}

The present study establishes a remarkable property of the SC states:  the handy, albeit obscure, negativity \cite{vidal} measure of  entanglement admits a clear geometrical interpretation. It is found that the negativity equals half the sum of the absolute values of the off-diagonal elements of the density matrix, which is a distance of the SC state from a pertinent separable state (see Fig. (\ref{one})). As a consequence, unlike a general mixed state \cite{peres96, Horodeckii}, the a SC state is separable \cite{werner} if and only if its negativity vanishes. It should be noted, that the matrix norm used to define the distance, permits an estimation of the negativity "at a glance", which has a strong intuitive appeal. 
Quantum-dynamical considerations motivate a generalization of the results to particular mixtures of SC states. It is shown that the negativity of such a mixture is less or equal to the distance of the  state from a pertinent separable state. 

We start from a formal definition of a SC state:

\textbf{Definition}. \textit{A mixed bipartite state $\Op \rho=\sum_i p_i \left|\psi_i\right\rangle\left\langle \psi_i\right|$ is called Schmidt-correlated if 
$\left|\psi_i\right\rangle=\sum_m c_m^i \left|m\right\rangle_1\otimes\left|m\right\rangle_2$ for every $i$, i.e. all pure states in the mixture share the same Schmidt bases $\Gamma_1=\left\{\left|m \right\rangle_1\right\}_{m=1}^N$ and $\Gamma_2=\left\{\left|m\right\rangle_2\right\}_{m=1}^N$.}

\textbf{{Theorem 1}}.\textit{ Let a bipartite state $\Op \rho$ be a SC state with respect to  Schmidt bases $\Gamma_1$ and $\Gamma_2$. Then the negativity of the state  $\Op \rho$ equals a distance of $\Op \rho$ from a  separable state $\Op \rho'$ diagonal in the tensor-product basis $\Gamma=\Gamma_1 \otimes \Gamma_2$: ${\cal N}(\Op \rho)=\frac{1}{2}d(\Op \rho,\Op \rho')$, where $ (\Op  \rho')_{ij}=\delta_{ij} (\Op  \rho)_{ij}$, $i, j\in \Gamma$ and the distance $d(\Op x,\Op y)=\left\|\Op x-\Op y\right\|_{\alpha}$ is induced by the matrix norm $\left\|\Op x\right\|_{\alpha}=\sum_{i,j}|(\Op x)_{i,j}| $} \cite{Bellman} .

\textit{Proof:} Let $\Op \rho=\sum_i p_i \left|\psi_i\right\rangle\left\langle \psi_i\right|$, where $\left|\psi_i\right\rangle=\sum_{m=1}^N c_m^i \left|m\right\rangle_1\otimes\left|m\right\rangle_2$. Then $\Op \rho=\sum_i p_i \sum_{mn}c_n^i(c_m^i)^*\left|n \right\rangle \left\langle m \right|\otimes \left| n\right\rangle \left\langle m\right|= \sum_{mn}\rho_{mn}\left|n \right\rangle \left\langle m \right|\otimes \left| n\right\rangle \left\langle m\right|$, where $\rho_{mn}=\sum_i p_i c_n^i(c_m^i)^*$. By definition $d(\Op \rho,\Op \rho')=\left\|\Op \rho-\Op \rho'\right\|_{\alpha}=\sum_{mn}|\rho_{mn}|-1=2\sum_{m<n}|\rho_{mn}|$. 

The negativity of a state is defined as the absolute value of the sum of the negative eigenvalues of the partially transposed density operator corresponding to the state \cite{vidal}. In what follows we show that $\left\{-|\rho_{mn}| , \ m<n\right\}$ is the set of all the negative eigenvalues of  the partially transposed $\Op \rho$. This completes the proof.

The partially transposed density matrix is given by 
\begin{eqnarray}
 \Op \rho^{PT}= \sum_{mn}\rho_{mn}\left|n \right\rangle \left\langle m \right|\otimes (\left| n\right\rangle \left\langle m\right|)^T=\sum_{mn}\rho_{mn}\left|n \right\rangle \left\langle m \right|\otimes \left| m\right\rangle \left\langle n\right|.
\end{eqnarray}
Consider $N(N-1)$ vectors 
\begin{eqnarray}
 \left|\psi_{kl}\right\rangle_{\pm}&=& - \rho_{lk}\left|k\right\rangle\left|l\right\rangle\pm|\rho_{kl}|\left|l\right\rangle\left|k \right\rangle , \ k < l, k=1,2,...,N-1
\end{eqnarray}
 and $N$ vectors 
\begin{eqnarray}
 \left|\psi_{kk}\right\rangle &=& \left|k\right\rangle\left|k\right\rangle, k=1,2,...,N.
\end{eqnarray}

\begin{eqnarray}
  \Op \rho^{PT}\left|\psi_{kl}\right\rangle_{\pm}&=& \sum_{mn}\rho_{mn}\left|n \right\rangle \left\langle m \right|\otimes \left| m\right\rangle \left\langle n\right| (- \rho_{lk}\left|k\right\rangle\left|l\right\rangle\pm|\rho_{kl}|\left|l\right\rangle\left|k \right\rangle ) \nonumber \\
  &=&- \sum_{mn}\rho_{mn}  \rho_{lk} \delta_{mk}\delta_{nl}\left|n \right\rangle  \left| m\right\rangle \pm \sum_{mn}\rho_{mn}|\rho_{kl}|\delta_{ml}\delta_{kn}\left|n \right\rangle  \left| m\right\rangle \nonumber \\
  &=&- \rho_{kl}  \rho_{lk} \left|l \right\rangle  \left| k\right\rangle \pm \rho_{lk}|\rho_{kl}|\left|k \right\rangle  \left| l\right\rangle=- |\rho_{lk}|^2 \left|l \right\rangle  \left| k\right\rangle \pm \rho_{lk}|\rho_{kl}|\left|k \right\rangle  \left| l\right\rangle \nonumber \\
  &=&\mp |\rho_{kl}|(- \rho_{lk}\left|k\right\rangle\left|l\right\rangle\pm|\rho_{kl}|\left|l\right\rangle\left|k \right\rangle)=\mp |\rho_{kl}|\left|\psi_{kl}\right\rangle_{\pm}.
\end{eqnarray}
Analogously we obtain
\begin{eqnarray}
  \Op \rho^{PT}\left|\psi_{kk}\right\rangle&=& |\rho_{kk}|\left|\psi_{kk}\right\rangle.
\end{eqnarray}
Thus  the partially transposed matrix $\Op \rho^{PT}$ has been diagonalized and  its $\frac{N(N-1)}{2}$ negative eigenvalues $-|\rho_{kl}|$, $k<l$ have been found , which completes the proof.
$\Box$ 

\textbf{Corollary 1}. \textit{Let a bipartite state $\Op \rho$ be  SC. Then $\Op \rho$
is disentangled if and only if its negativity vanishes.}

\textit{Proof:} If $\Op \rho$
is disentangled its negativity ${\cal N}( \Op \rho)$ vanishes by Peres-Horodeckii criterion \cite{peres96, Horodeckii}. 
If ${\cal N}( \Op \rho)=0$ then by Theorem 1 $\Op \rho$ is separable.
$\Box$

The following two corollaries  of Theorem 1 permit an estimation of entanglement of a SC state by simply "looking at"  the occupied entries of the corresponding density matrix. 

\textbf{Corollary 2}.\textit{ The negativity of the SC (with respect to Schmidt bases $\Gamma_{1,2}$) state $\Op \rho$ equals half the sum of the off-diagonal elements of the corresponding density matrix in the Schmidt basis: ${\cal N}(\Op \rho)=\frac{1}{2}\sum_{i\neq j}|\rho_{ij}|$, $i,j\in \Gamma_1\otimes\Gamma_2$. }

\textbf{Corollary 3}. \textit{Let a  SC state $\Op \rho$ be quasidiagonal, i.e  $\Op \rho= \sum_{mn}\rho_{mn}\left|n \right\rangle \left\langle m \right|\otimes \left| n\right\rangle \left\langle m\right|$, where $m,n =1,2,...,N$ and $|m-n|=\Delta < N-1$. Then ${\cal N}(\Op \rho)< \Delta$.}

\textit{Proof:} By Corollary 2 it suffices to show that $\frac{1}{2}\sum_{m\neq n}|\rho_{mn}| < \Delta$. The sum of the absolute values of the off-diagonal elements can be estimated as follows:
 \begin{eqnarray}
 \sum_{m\neq n}|\rho_{mn}|&=&  \sum_{m n}|\rho_{mn}|-1 = \sum_{m}\sum_{n=m-\Delta}^{n=m+\Delta}|\rho_{mn}|-1 <  \sum_{m}\sum_{n=m-\Delta}^{n=m+\Delta}\sqrt{\rho_{mm}\rho_{nn}} \nonumber \\ \\
 &\le &\sum_{m}\sum_{n=m-\Delta}^{n=m+\Delta}\frac{\rho_{mm}+\rho_{nn}}{2}=\frac{1}{2}\sum_{m}\sum_{n=m-\Delta}^{n=m+\Delta}\rho_{mm}+\frac{1}{2}\sum_{m}\sum_{n=m-\Delta}^{n=m+\Delta}\rho_{nn} < 2\Delta, \nonumber
\label{eq:estimation}
\end{eqnarray}
where the first inequality follows from the positivity of the density operator and the second is the inequality of geometric and arithmetic means. This concludes the proof.
$\Box$

The SC correlated states naturally emerge in certain quantum dynamical settings (cf. Ref.\cite{Khasin06}).
Assume a (generally non unitary) evolution of a bipartite composite system admitting an additive integral of motion $\Op A=\Op A_1\otimes \Op I_2+\Op I_1\otimes \Op A_2$, i.e.
\begin{eqnarray}
  \frac{\partial}{\partial t}\Op \rho= {\cal L }\Op \rho
\label{eq:liouv}
\end{eqnarray}
and
\begin{eqnarray}
  \frac{d}{d t}\Op A= {\cal L }^{\dagger}\Op A=0.
\end{eqnarray}
Consider  local  bases of eigenstates of operators ${\Op A}_{1}$ and ${\Op A}_{2}$:
\begin{eqnarray}
  \Gamma_{i}= \left\{ \left|m\right\rangle_{i}\ m=1,2,...,N_{i}, \ \Op A_i\left|m\right\rangle_i=\lambda_m^i \left|m\right\rangle_i \right\}, \ i=1,2.
  \label{eq:bases}
\end{eqnarray}
We denote as ${\cal H}^{\lambda}$ a subspace  of the composite system Hilbert space ${\cal H}$ spanned by the eigenstates of $\Op A$, corresponding to an eigenvalue $\lambda$:
\begin{eqnarray}
  {\cal H}^{\lambda}=\texttt{Sp}\left\{ \left|m\right\rangle_1\otimes\left|n\right\rangle_2, \lambda_m^1+\lambda_n^2=\lambda \right\}.
\end{eqnarray}
Let us assume that the spectra of $\Op A_i$, $i=1,2$ are non degenerate, i.e. $\lambda_m^i=\lambda_n^i$ $\Rightarrow$ $m=n$. Then the equation $\lambda_m^1+\lambda_n^2=\lambda$  with fixed $m$ and $\lambda$ possesses a unique solution for $n$: $n=f^{\lambda}(m)$. Therefore the map $f^{\lambda}: \Gamma_1 \rightarrow \Gamma_2$ 
provides a one-to-one correspondence between a state $\left|m\right\rangle_{1}\in \Gamma_1$ and a state $\left|m\right\rangle_{2}\in \Gamma_2$, i.e. it defines  a unique common set of Schmidt bases for the Schmidt decomposition of all $\left|\psi^{\lambda} \right\rangle\in {\cal H}^{\lambda}$:
\begin{eqnarray}
  \left|\psi^{\lambda}\right\rangle=\sum_{\lambda_m^1+\lambda_n^2=\lambda}c_m \left|m\right\rangle_1\left|n\right\rangle_2.
\end{eqnarray}
Since all pure states $\left|\psi^{\lambda}\right\rangle\in {\cal H}^{\lambda}$ share the same Schmidt  bases  their mixture  is a SC  state.

If the initial state of the composite system is a mixture of eigenstates of $\Op A$, corresponding to the same eigenvalue $\lambda$ , i.e. $\Op \rho(0)=\sum_i p_i \left|\psi^{\lambda}_i\right\rangle\left\langle \psi^{\lambda}_i\right|$ with $\left|\psi^{\lambda}_i\right\rangle\in {\cal H}^{\lambda}$ then the conservation of $\Op A$ would imply that $\Op \rho(t)=\sum_i p_i(t) \left|\psi_i(t)\right\rangle\left\langle \psi_i(t)\right|$,  $\left|\psi(t)\right\rangle_i\in {\cal H}^{\lambda}$  at any $t>0$. Therefore, the evolving state $\Op \rho(t)$ remains SC and the negativity of the state can be calculated using Theorem $1$. 

 For illustration, the evolution of negativity of a composite state of a  two noninteracting quantum systems coupled to a local purely dephasing baths is calculated. The composite system evolves according to the Liouville equation
\begin{eqnarray}
  \frac{\partial}{\partial t}\Op \rho= -[\Op A_1,[\Op A_1,\Op \rho]]-[\Op A_2,[\Op A_2,\Op \rho]].
\label{eq:liouvloc}
\end{eqnarray}
The local  bases of eigenstates of operators ${\Op A}_{1}$ and ${\Op A}_{2}$ is $\Gamma_{i}= \left\{ \left|m\right\rangle_{i}, \ m=1,2,...,N_{i}, \ \Op A_i \left|m\right\rangle_i=\lambda_m^i\left|m\right\rangle_i \right\}, \ i=1,2.$
The initial  state of the composite system is a pure entangled state
\begin{eqnarray}
  \left|\psi(0)\right\rangle=\sum_m c_m\left|m\right\rangle_1\left|m\right\rangle_2.
\label{eq:initial}
\end{eqnarray}
 At $t>0$ the solution of Eq.(\ref{eq:liouvloc}) with initial state (\ref{eq:initial}) is given by 
\begin{eqnarray}
 \Op \rho(t)=\sum_{m,n} c_m c_n^* e^{-\left[(\lambda_m^1-\lambda_n^1)^2+(\lambda_m^2-\lambda_n^2)^2\right]t}\left|m\right\rangle_1\left|m\right\rangle_2\left\langle n\right|_1\left\langle n\right|_2.
\end{eqnarray}
Since $\Op \rho(t)$ is a SC state the Theorem 1 applies and 
\begin{eqnarray}
 {\cal N}(\Op \rho(t))&=& \frac{1}{2}d(\Op \rho(t),\Op \rho'(t))= \frac{1}{2} \left(\sum_{i,j\in \Gamma_1\otimes \Gamma_2 }|(\Op \rho^{\lambda})_{i,j}|-1\right) \nonumber \\
 &=& \frac{1}{2} \left(\sum_{m,n} c_m c_n^* e^{-\left[(\lambda_m^1-\lambda_n^1)^2+(\lambda_m^2-\lambda_n^2)^2\right]t}-1\right).
\end{eqnarray}

As a next step, the initial state for the evolution (\ref{eq:liouv}) is generalized from a state $\Op \rho(0)=\sum_i p_i \left|\psi^{\lambda}_i\right\rangle\left\langle \psi^{\lambda}_i\right|$ with $\left|\psi^{\lambda}_i\right\rangle\in {\cal H}^{\lambda}$ to a mixture of such states, corresponding to different eigenvalues $\lambda$ of the conserved operator $\Op A$:
 \begin{eqnarray}
 \Op \rho(0)=\sum_{\lambda} p_{\lambda} \Op \rho^{\lambda},
\end{eqnarray}
where $\Op \rho^{\lambda}=\sum_i p_i^{\lambda} \left|\psi^{\lambda}_i\right\rangle\left\langle \psi^{\lambda}_i\right|$ with $\left|\psi^{\lambda}_i\right\rangle\in {\cal H}^{\lambda}$.
By conservation of $\Op A$ we have $\Op \rho(t)=\sum_{\lambda} p_{\lambda}(t) \Op \rho^{\lambda}(t)$. An estimation of the negativity of $\Op \rho(t)$ is possible using a generalization of the Theorem 1 (see Theorem 2 below).  The result is:
\begin{eqnarray}
  {\cal N}(\Op \rho(t))\le \frac{1}{2}d(\Op \rho(t),\Op \rho'(t)),
\end{eqnarray}
 where  $ (\Op  \rho'(t))_{ij}=\delta_{ij} (\Op  \rho(t))_{ij}$, $i,j\in \Gamma_1\otimes \Gamma_2$ (see Eq.(\ref{eq:bases})) and the distance $d(\Op x,\Op y)=\left\|\Op x-\Op y\right\|_{\alpha}$ is induced by the norm $\left\|\Op x\right\|_{\alpha}=\sum_{i,j}|(\Op x)_{i,j}|$.

Theorem 2 generalizes Theorem 1 to particular \textit{mixtures} of SC states. 
Consider a composite Hilbert space ${\cal H}={\cal H}_1\otimes {\cal H}_2$ of bipartite quantum system. 
Let $\Gamma_{1}=\left\{\left|m\right\rangle_{1}, m=1,2,...,N_1\right\}$ be an orthonormal basis of the local Hilbert space ${\cal H}_{1}$ and $\Gamma_{2}=\left\{\left|m\right\rangle_{2}, m=1,2,...,N_2\right\}$ be an orthonormal basis of the local Hilbert space ${\cal H}_{2}$. Consider a one-to-one correspondence $f^{\lambda}$ between a subset $S^{\lambda}_1\subset \Gamma_1$ to a subset $S^{\lambda}_2\subset \Gamma_2$. The map $f^{\lambda}$ defines a subspace ${\cal H}^{\lambda}\subset {\cal H}$ spanned by the states of the form $\left|m \right\rangle_1 \otimes \left|f^{\lambda}(m)\right\rangle_2$: 
\begin{eqnarray}
 {\cal H}^{\lambda}= \texttt{Sp}\left\{ \left|m \right\rangle_1 \otimes \left|f^{\lambda}(m)\right\rangle_2, \ \left|m \right\rangle_1\in  S^{\lambda}_1 \right\}.
\end{eqnarray}
All pure states $\left|\psi^{\lambda}\right\rangle \in {\cal H}^{\lambda}$ share the same Schmidt bases by construction. Therefore, a mixture $\Op \rho^{\lambda}=\sum_i p_i \left|\psi^{\lambda}_i\right\rangle\left\langle \psi^{\lambda}_i\right|$ is a SC state by definition and will be called \textit{$\lambda$-SC state} for brevity in what follows. 

Let us consider a family of one-to-one maps $f^{\lambda}$ from $\Gamma_1$ to $\Gamma_2$ with a property:
\begin{eqnarray}
 f^{\lambda_1}(\left|m\right\rangle_{1})=f^{\lambda_2}(\left|m\right\rangle_{1})   \Leftrightarrow \lambda_1=\lambda_2.
 \label{eq:disjoint}
\end{eqnarray}
  Then the following result can be proved.

\textbf{{Theorem 2}}.\textit{ Let a bipartite state $\Op \rho$ be a mixture of $\lambda$-SC states: $\Op \rho=\sum_{\lambda} p_{\lambda} \Op \rho^{\lambda}$ with respect to local bases $\Gamma_{1,2}$ and a family of maps $f^{\lambda}: \Gamma_1 \rightarrow \Gamma_2$ with the property (\ref{eq:disjoint}). Then the negativity of the state  $\Op \rho$ is less or equals a distance of $\Op \rho$ from a  separable state $\Op \rho'$ diagonal in the basis $\Gamma=\Gamma_1 \otimes \Gamma_2$: ${\cal N}(\Op \rho)\le \frac{1}{2}d(\Op \rho,\Op \rho')$, where $ (\Op  \rho')_{ij}=\delta_{ij} (\Op  \rho)_{ij}$, $i,j\in \Gamma$ and the distance $d(\Op x,\Op y)=\left\|\Op x-\Op y\right\|_{\alpha}$, where $\left\|\Op x\right\|_{\alpha}=\sum_{i,j}|(\Op x)_{i,j}|$.}

\textit{Proof:} Since the negativity is entanglement monotone \cite{vidal} the following holds:
\begin{eqnarray}
  {\cal N}( \sum_{\lambda} p_{\lambda} \Op \rho^{\lambda})\le \sum_{\lambda} p_{\lambda}  {\cal N}( \Op \rho^{\lambda}).
\label{eq:mono}
\end{eqnarray}
By Theorem 1:
\begin{eqnarray}
\sum_{\lambda} p_{\lambda}  {\cal N}( \Op \rho^{\lambda})&=&\sum_{\lambda} p_{\lambda}  \frac{1}{2}d(\Op \rho^{\lambda},\Op \rho'^{\lambda})=\sum_{\lambda} p_{\lambda}  \frac{1}{2}\left\|\Op \rho^{\lambda}-\Op \rho'^{\lambda}\right\|_{\alpha}\nonumber \\
&=&\sum_{\lambda} p_{\lambda}  \frac{1}{2}\sum_{i,j\in \Gamma}|(\Op \rho^{\lambda}-\Op \rho'^{\lambda})_{i,j}|=\sum_{\lambda} p_{\lambda}  \frac{1}{2}\left(\sum_{i,j\in \Gamma}|(\Op \rho^{\lambda})_{i,j}|-1\right)\nonumber \\
&=& \frac{1}{2} \left(\sum_{\lambda} p_{\lambda}  \sum_{i,j\in \Gamma}|(\Op \rho^{\lambda})_{i,j}|-1\right).
\label{eq:nine}
\end{eqnarray}
From the property (\ref{eq:disjoint}) it follows that 
\begin{eqnarray}
\sum_{\lambda} p_{\lambda}  \sum_{i,j\in \Gamma}|(\Op \rho^{\lambda})_{i,j}|=\sum_{i,j\in \Gamma} |\left(\sum_{\lambda} p_{\lambda}\Op \rho^{\lambda}\right)_{i,j}|=\sum_{i,j\in \Gamma} |({\Op \rho})_{i,j}|.
\label{eq:ten}
\end{eqnarray}
Combining Eqs. (\ref{eq:nine}) and (\ref{eq:ten}) we get
\begin{eqnarray}
\sum_{\lambda} p_{\lambda}  {\cal N}( \Op \rho^{\lambda})&=&\frac{1}{2} \left(\sum_{\lambda} p_{\lambda}  \sum_{i,j\in \Gamma}|(\Op \rho^{\lambda})_{i,j}|-1\right)=\frac{1}{2} \left(\sum_{i,j\in \Gamma} |({\Op \rho})_{i,j}|-1\right)\nonumber \\
&=&\frac{1}{2} d(\Op \rho,\Op \rho'),
\label{eq:final}
\end{eqnarray}
 From inequality (\ref{eq:mono}) and Eq.(\ref{eq:final}) it follows that
\begin{eqnarray}
{\cal N}(\Op \rho)\le \frac{1}{2}d(\Op \rho,\Op \rho').
\end{eqnarray}
$\Box$ 

Corollaries 1-3 of the Theorem 1 can be  generalized accordingly.

In conclusion, the negativity of a SC state can be interpreted geometrically as a
distance in a particular metric $d$ of the state from a separable state. An immediate consequence of this fact is that the negativity vanishes if and only if the SC state is separable. The metric $d$ that is induced by the  $\alpha$ matrix norm, is  basis dependent, i.e. is not invariant under unitary transformations (and, in particular, is not invariant under local unitary transformations). Nevertheless, the basis in which the correspondence of this distance to the negativity is established is the Schmidt basis, which is a preferred basis for representing SC states. 

In a SC state the negativity equals half the sum of the absolute values of the off-
diagonal elements of the density matrix in the Schmidt bases. This finding
suggests the "at a glance" estimation of the entanglement of SC states: the state is "substantially" entangled if and only if the off-diagonal entries in the corresponding
density matrix are "substantially" populated. In particular, if the corresponding density matrix is quasidiagonal, i.e. the off-diagonal elements populate the strip about the diagonal of width  $\Delta$, the negativity is bounded by $\Delta$.

We have considered Schmidt-correlated states and particular mixtures of Schmidt-correlated states.  
These states emerge in dynamical models with conservation laws. Dynamics where the conservation laws are relaxed generate mixed states that are not SC.  Simulations  of open-system dynamics, similar to those in Ref.\cite{Khasin06}, have  suggested a generalization of the geometrical
interpretation of negativity to arbitrary mixed states. It is conjectured that
the negativity of an arbitrary mixed state is bounded by the minimal distance $d$
of the state to a corresponding separable state, where the distance is minimized
over all possible local bases.
From this conjecture it follows  that the negativity of an arbitrary state is bounded by half the sum of the off-diagonal elements of the corresponding density matrix in \textit{any local bases}, which gives an intuitive appraisal of the negativity (and entanglement) of an arbitrary mixed state.

 \end{document}